\documentclass[letterpaper,english]{emulateapj}
\usepackage{ae,aecompl}
\usepackage[T1]{fontenc}
\usepackage[latin9]{inputenc}
\usepackage{array}
\usepackage{booktabs}
\usepackage{multirow}
\usepackage{amstext}
\usepackage{graphicx}
\PassOptionsToPackage{normalem}{ulem}
\usepackage{ulem}

\makeatletter

\pdfpageheight\paperheight
\pdfpagewidth\paperwidth

\providecommand{\tabularnewline}{\\}

\slugcomment{}

\shorttitle{}

\shortauthors{}

\usepackage{enumitem}  

\setlist{nolistsep}

\usepackage{graphicx}

\usepackage{placeins}

\usepackage{amsmath}

\makeatother

\usepackage{babel}
\begin{document}

\title{Spitzer Observations Confirm and Rescue the Habitable-Zone Super-Earth
K2-18b for Future Characterization}

\author{Björn Benneke$^{1}$, Michael Werner$^{2}$, Erik Petigura$^{1}$,
Heather Knutson$^{1}$, Courtney Dressing$^{1,9}$, Ian J. M. Crossfield$^{3,8,9}$,
Joshua E. Schlieder$^{4}$, John Livingston$^{2}$, Charles Beichman$^{5}$,
Jessie Christiansen$^{6}$, Jessica Krick$^{6}$, Varoujan Gorjian$^{2}$,
Andrew W. Howard$^{7}$, Evan Sinukoff$^{7}$, David R. Ciardi$^{5}$,
Rachel L. Akeson$^{6}$}

\affil{$^{1}$Division of Geological and Planetary Sciences, California
Institute of Technology, Pasadena, CA 91125, USA}

\affil{$^{2}$Jet Propulsion Laboratory, California Institute of Technology}

\affil{$^{3}$Lunar \& Planetary Laboratory, University of Arizona, 1629
E. University Blvd., Tucson, AZ, USA}

\affil{$^{4}$NASA Ames Research Center, Moffett Field, CA, USA}

\affil{$^{5}$NASA Exoplanet Science Institute, California Institute of
Technology, 770 S. Wilson Ave., Pasadena, CA, USA}

\affil{$^{6}$Infrared Processing and Analysis Center, California Institute
of Technology, Pasadena, CA 91125, USA}

\affil{$^{7}$Institute for Astronomy, University of Hawaii, 2680 Woodlawn
Drive, Honolulu, HI, USA}

\affil{$^{8}$Department of Astronomy and Astrophysics, University of California,
Santa Cruz, CA 95064, USA}

\affil{$^{9}$NASA Sagan Fellow}

\email{bbenneke@caltech.edu}
\begin{abstract}
The recent detections of two transit events attributed to the super-Earth
candidate K2-18b have provided the unprecedented prospect of spectroscopically
studying a habitable-zone planet outside the Solar System. Orbiting
a nearby M2.5 dwarf and receiving virtually the same stellar insolation
as Earth, K2-18b would be a prime candidate for the first detailed
atmospheric characterization\textit{ }of a habitable-zone exoplanet
using \textit{HST} and \textit{JWST.} Here, we report the detection
of a third transit of K2-18b near the predicted transit time using
the \textit{Spitzer Space Telescope}. The \textit{Spitzer} detection
demonstrates the periodic nature of the two transit events discovered
by \textit{K2}, confirming that K2-18 is indeed orbited by a super-Earth
in a 33-day orbit and ruling out the alternative scenario of two similarly-sized,
long-period planets transiting only once within the 75-day \textit{K2}
observation. We also find, however, that the transit event detected
by \textit{Spitzer} occurred 1.85 hours ($7\,\sigma$) before the
predicted transit time. Our joint analysis of the \textit{Spitzer}
and \textit{K2} photometry reveals that this early occurrence of the
transit is not caused by transit timing variations (TTVs), but the
result of an inaccurate \textit{K2} ephemeris due to a previously
undetected data anomaly in the\textit{ K2} photometry likely caused
by a cosmic ray hit. We refit the ephemeris and find that K2-18b would
have been lost for future atmospheric characterizations with \textit{HST}
and \textit{JWST} if we had not secured its ephemeris shortly after
the discovery. We caution that immediate follow-up observations as
presented here will also be critical in confirming and securing future
planets discovered by \textit{TESS}, in particular if only two transit
events are covered by the relatively short 27-day \textit{TESS} campaigns.
\end{abstract}

\section{Introduction\label{sec:Introduction}}

Ever since the discovery of the first planet orbiting a Sun-like star,
we have made steady progress towards finding habitable zone exoplanets
with the eventual goal to characterize their atmospheres and climates
in great detail. Results from the \textit{Kepler} mission, which was
the first telescope with the sensitivity to detect small planets in
the habitable zones of sun-like stars, indicate that the occurrence
rate for habitable Earths (\textquotedblleft $\eta_{\oplus}$\textquotedblright )
may be as high as 5-20\% \citep{petigura_plateau_2013,foreman-mackey_exoplanet_2014,silburt_statistical_2015,farr_counting_2015,burke_terrestrial_2015}.
The majority of these planets found from the \textit{Kepler} mission,
however, are orbiting distant and faint stars, preventing us from
measuring their bulk masses or atmospheric compositions through spectroscopy. 

The recent announcement of the super-Earth candidate K2-18b \citep{montet_stellar_2015}
orbiting in the habitable zone of a nearby bright M2.8-dwarf provides
the unprecedented opportunity to characterize the first atmosphere
of a habitable zone planet outside our solar system. Atmospheric studies
of K2-18b are within the reach of currently available instrumentation
because the radius of the host star K2-18 is only $0.39\, R_{\odot}$
and the planet\textquoteright s radius of $2.2\, R_{\oplus}$ allows
for the presence of a hydrogen-rich, low mean molecular weight gas
envelope. Water and methane absorption in such a hydrogen-rich atmosphere
could result in transit depth variations that can be revealed through
\textit{HST}, \textit{Spitzer}, and/or \textit{JWST} transit observations.

The two transit events of the planet candidate K2-18b were originally
discovered by analyzing the Campaign 1 data from the extended \textit{Kepler}
(\textquotedblleft \textit{K2}\textquotedblright ) mission \citep{foreman-mackey_systematic_2015}.
Modern seeing-limited images and adaptive optics imaging subsequently
ruled out background eclipsing binaries as a possible source for the
detected transit events \citep{montet_stellar_2015}. Radial velocity
measurements further eliminated the possibility that the apparent
transit events were caused by non-planetary companions co-moving with
K2-18b.

\begin{figure}[t]
\noindent \centering{}\includegraphics[width=1\columnwidth]{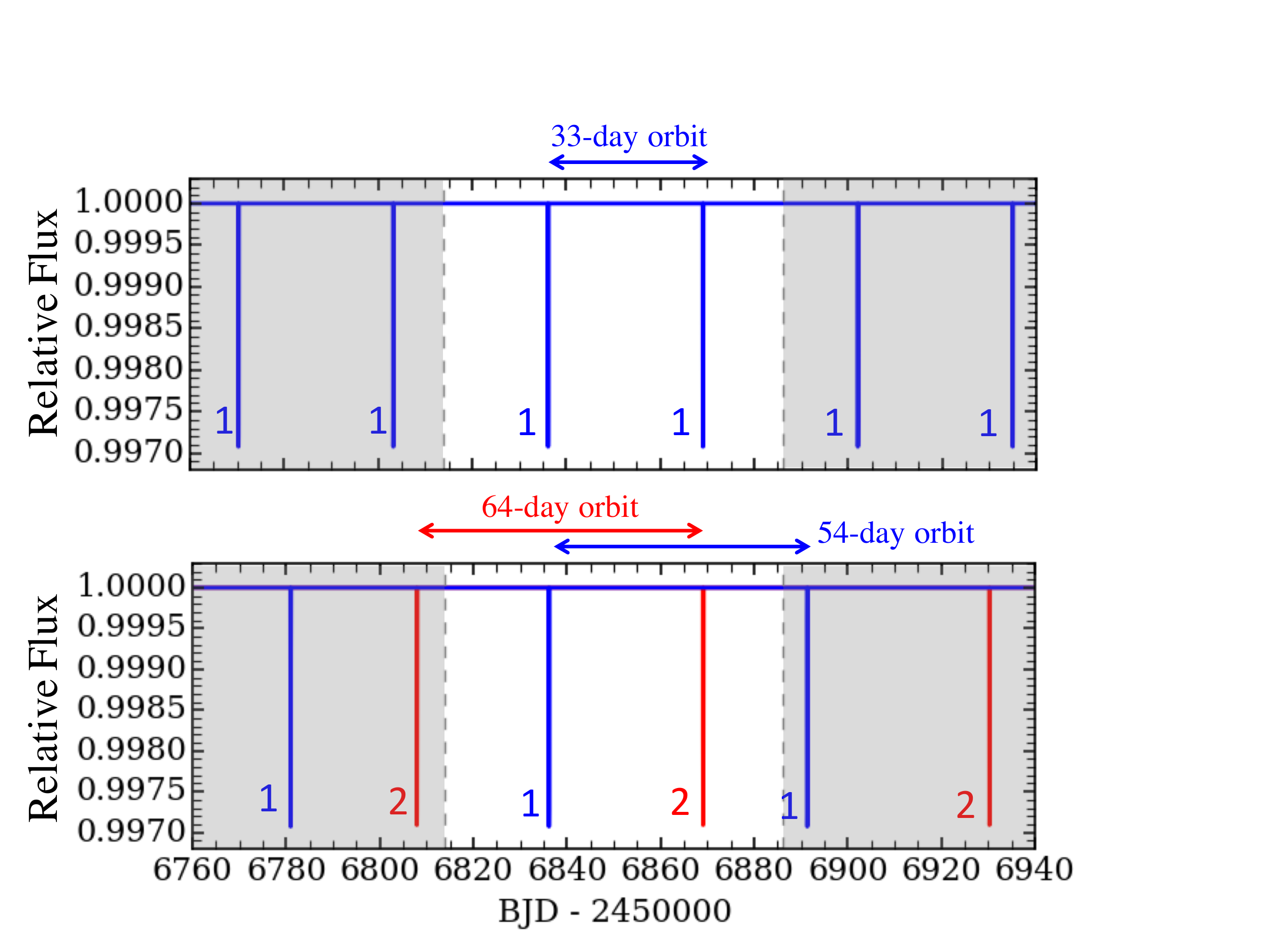}\protect\caption{Plausible scenarios in agreement with the available \textit{K2} data
of K2-18. The white region indicates the time span observed by \textit{K2}.
The two detected transit events at $\mathrm{BJD=2,456,836}$ and $\mathrm{BJD=2,456,869}$
can be fit equally well by a single planet in a 33-day orbit at low
impact parameter (top panel) or two long-period planets at slightly
higher impact parameter (bottom panel).\label{fig:Plausible-scenarios-in}}
\end{figure}

The \textit{K2} photometry, however, could not demonstrate the periodic
nature of the transit signal. Instead, the 80-day \textit{K2} photometry
identified only two transit events of similar depth with a separation
of 33-days. Meanwhile, radial velocity confirmation of the 33-day
period is currently not available because the host star is faint at
visible wavelengths ($V=13.5)$ and the expected semi-amplitude of
a 33-day-period planet would only be $K_{S}=1-2\,\mbox{m/s}$. As
a result, the available observations leave open the alternative scenario
that the two detected transit events on 2014-June-27 and 2014-July-30
were caused by two similarly-sized, long-period planets ($\gtrsim50$~days)
that happen to transit only once within the 80 days observation in
\textit{K2} Campaign 1 (Figure \ref{fig:Plausible-scenarios-in}).
In this two-planet scenario, a planet with a 33-day orbital period
and Earth-like incident stellar irradiance ($S_{inc}\approx S_{\oplus}$)
would not exist around the star K2-18. We find that two-planet scenarios
can lead to equally good fits to the long-cadence \textit{K2} data
because the increase in transit duration with orbital period can effectively
be compensated by a high impact parameter and/or an eccentric orbit. 

To distinguish between the one and two planet scenario, we obtained
\textit{Spitzer }high precision photometry at $4.5\,\mu m$ to probe
for a third transit event that would only occur if a habitable-zone
super-Earth indeed existed in a 33-day orbit around K2-18. A detection
of a third transit at the predicted time would prove the periodic
nature of the signal. It would simultaneously also rule out potential
scenarios in which one or both of the identified transit-like events
were the result of residual systematic effects in the corrected \textit{K2}
photometry. Such scenarios are conceivable because the detected signals
are well below the noise floor of the uncorrected \textit{K2} photometry
and outliers well above the median light curve scatter are common
in the corrected \textit{K2} photometry despite state-of-the-art detrending
\citep[e.g.,][]{vanderburg_technique_2014,foreman-mackey_systematic_2015}.

The article is structured as follows: In Section 2, we describe the
\textit{Spitzer} and \textit{K2} observations used in this work as
well as our photometric extraction routines. Section 3 presents the
light curves analyses. Finally, we discuss the results and conclusions
in Section 4.

\begin{figure}[t]
\noindent \centering{}\includegraphics[width=1\columnwidth]{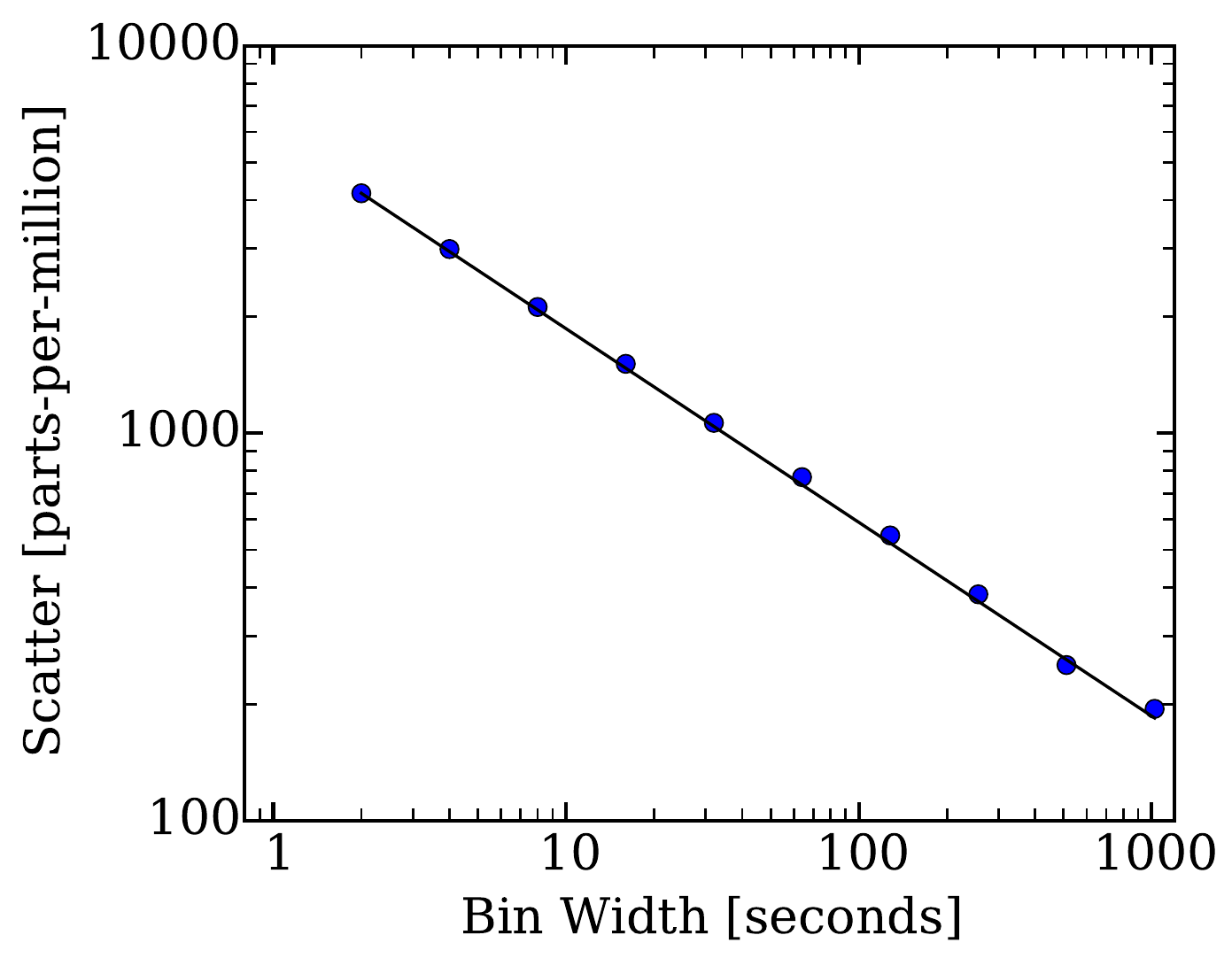}\protect\caption{Photometric scatter versus the width of the binning interval for \textit{Spitzer}
data of K2-18b. The root-mean-square error of the systematics-corrected
\textit{Spitzer} data (blue points) follows closely the theoretical
square-root scaling for uncorrelated white noise (black line). The
left-most data point corresponds to the unbinned 2-second exposures.
\label{fig:RedNoise}}
\end{figure}

\begin{figure}[t]
\noindent \centering{}\includegraphics[width=1\columnwidth]{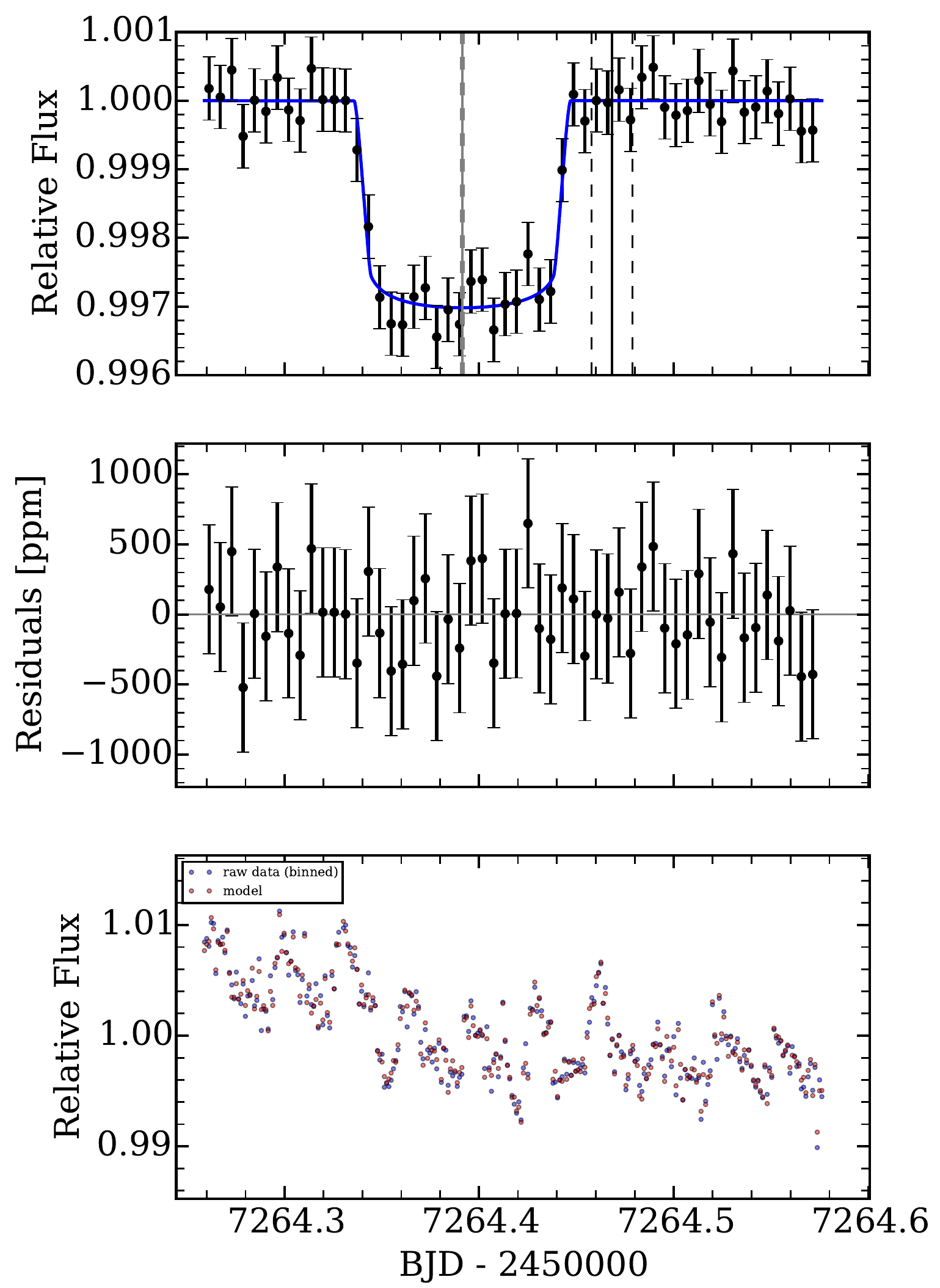}\protect\caption{Light curve fit to the \textit{Spitzer} transit observation of K2-18b.
The top panel shows the best fitting model light curve (blue), overlaid
with the systematics-corrected \textit{Spitzer} data. The transit
of K2-18b appears $7\sigma$ (1.85 hours) before the predicted transit
time from Montet et al. 2015 (vertical black line). Residuals from
the light curve fitting are plotted in the middle panel, with uncertainty
bars corresponding to the fitted photometric scatter. The final systematics-corrected
photometry is near the Poisson noise limit and virtually free of red
noise and systematics. For clarity, \textit{Spitzer} data are binned
to 10-minute intervals. The bottom panel shows the raw \textit{Spitzer}
photometry (blue), overlaid with the model fit (red). Only Spitzer
observations are used in this fit.\label{fig:Spitzer-light-curve-1-1}}
\end{figure}

\section{Observations}

\subsection{\textit{Spitzer/IRAC} \label{sub:Spitzer/IRAC-Observations}}

We observed the star K2-18 (EPIC~201912552) for a total of 8.1 hours
on August 29, 2015 to search for the predicted transit of the super-Earth
candidate K2-18b as part of our \textit{K2} follow-up program (GO~11026,
PI Werner, see also \citealt{beichman_spitzer_2016-2}). The science
observation began three hours before the start of the predicted transit
of K2-18b and ended two hours after the end of the predicted transit
to account for transit ephemeris uncertainties and to provide adequate
baselines on either side of the transit. An additional 30-minute of
pre-observation (not used in the analysis) preceded the science observation
to mitigate the initial instrument drift in the science observations
resulting from telescope temperature changes after slewing from the
preceding target \citep{grillmair_pointing_2012}. Both pre-observation
and science observations were taken using \textit{Spitzer/IRAC} Channel
2 in stare mode. To enhance the accuracy in positioning the target
star K2-18 on the IRAC detector, the pre-observations were taken in
peak-up mode using the Pointing Calibration and Reference Sensor (PCRS)
as a positional reference.

We chose \textit{Spitzer/IRAC} Channel 2 ($4.5\,\mathrm{\mu m}$)
because the instrumental systematic due to intra-pixel sensitivity
variations are smaller for Channel 2 than for Channel 1 \citep{ingalls_intra-pixel_2012}.
Our exposure times were set to 2 seconds to optimize the integration
efficiency while comfortably remaining in the linear regime of the
IRAC detector. Subarray mode was used to reduce the readout overhead
and lower the data volume for the downlink. In total, our science
data are composed of 13,632 individual frames forming a 7.6-hour broadband
photometric time series of K2-18 at $4.5\,\mathrm{\mu m}$. 

We extract multiple photometric light curves from the science data
using a wide range of fixed and variable aperture sizes. The purpose
of extracting and comparing multiple photometric light curves is to
choose the aperture that provides the lowest residual scatter and
red-noise component. For each aperture, our extraction includes estimating
and subtracting the sky background, calculating the flux-weighted
centroid position of the star on the array, and then calculating the
total flux within circular apertures \citep{knutson_3.6_2012,lewis_orbital_2013,todorov_warm_2013,kammer_spitzer_2015}.
For the fixed apertures we consider radii between 2.0 and 5.0 pixels.
For the time varying apertures, we first calculate the scaling of
the noise pixel parameter $\beta=(\sum_{n}I_{n})^{2}/(\sum_{n}I_{n}^{2})$,
where $I_{n}$ is the measured intensity in the $n$-th pixel \citep{mighell_stellar_2005},
and then iteratively rescale the noise pixel aperture radius as $r=a\sqrt{\beta}+c,$
where we explore values between 0.6 and 1.2 for the scaling factors
$a$ and values between 0.6 and 1.2 for the constant c. We choose
this range for a and c because we noted in previous \textit{Spitzer}
work that the photometric scatter increases outside this range \citep{kammer_spitzer_2015}.
Finally, we pick the version of the \textit{Spitzer} photometry with
the lowest red-noise component. Our red-noise measure is the summed
square difference between the noise scaling obtained by successively
doubling the bin size and the theoretical square-root noise scaling
for a Poisson process (Figure \ref{fig:RedNoise}). In our case, the
version of the photometry with the lowest red-noise also results in
the lowest RMS. 

Our analysis is performed on the entire 7.6-hour science data since
there is no evidence for a residual ramp effect at the beginning of
the science data. We normalize the light curve by the median value
and bin the data to 60-second cadence. We find that moderate binning
to 60 seconds does not affect the information content of the photometry,
but provides more signal per data point allowing an improved correction
of the systematics (Section 3). We calculate $\mathrm{BJD_{UTC}}$
mid-exposure times using the information in the header of the BCD
files provided by the \textit{Spitzer} pipeline. We show the resulting
uncorrected light curve for the fixed aperture with a radius of 3.0
pixels in Figure \ref{fig:Spitzer-light-curve-1-1} (bottom panel).

\subsection{\textit{K2} Photometry}

The star K2-18 was observed by the \textit{Kepler Space Telescope}
as part of \textit{K2} Campaign~1 covering an 80-day time span between
November 14, 2014 and February 3, 2015. We extract the \textit{K2}
photometry directly from the \textit{Kepler} pixel data downloaded
from the Mikulski Archive for Space Telescopes (MAST). The star is
listed as EPIC-201912552. The full data set is a time series composed
of 3737 individual 16x16 pixel images centered on the star K2-18.
The individual images have a cadence of 29.4 minutes and were obtained
by co-adding 270 exposures (each 6.02~seconds plus 0.52~seconds
readout) onboard the Kepler spacecraft \citep{gilliland_initial_2010}.
We do not do any additional time binning during the photometric extraction
and light curve analysis of the \textit{K2} data.

\subsubsection{Photometric Extraction}

Our photometric extraction routine is outlined in \citet{crossfield_nearby_2015}
and \citet{petigura_two_2015}, following the approach introduced
by \citet{vanderburg_technique_2014}. In brief, during the continuous
80-day \textit{K2} observations, the stars drift across the CCD by
approximately 1 pixel every 6 hours due to the spacecraft\textquoteright s
pointing jitter. As the stars drift through pixel-phase, intra-pixel
sensitivity variations and errors in the flatfield cause the apparent
brightness of the target star to change. We detrend the apparent brightness
variations of our target using the telescope roll angle between the
target frame and an arbitrary reference observation. For each of the
two transits, we then extract transit light curves ranging from three
hours before the transit ingress to three hours after ingress, providing
sufficient baseline for the transit light curve analysis.

\subsubsection{A Cosmic Ray Detection Algorithm for K2 Photometry\label{sub:A-Cosmic-Ray}}

\begin{figure*}[t]
\noindent \centering{}\includegraphics[width=1\textwidth]{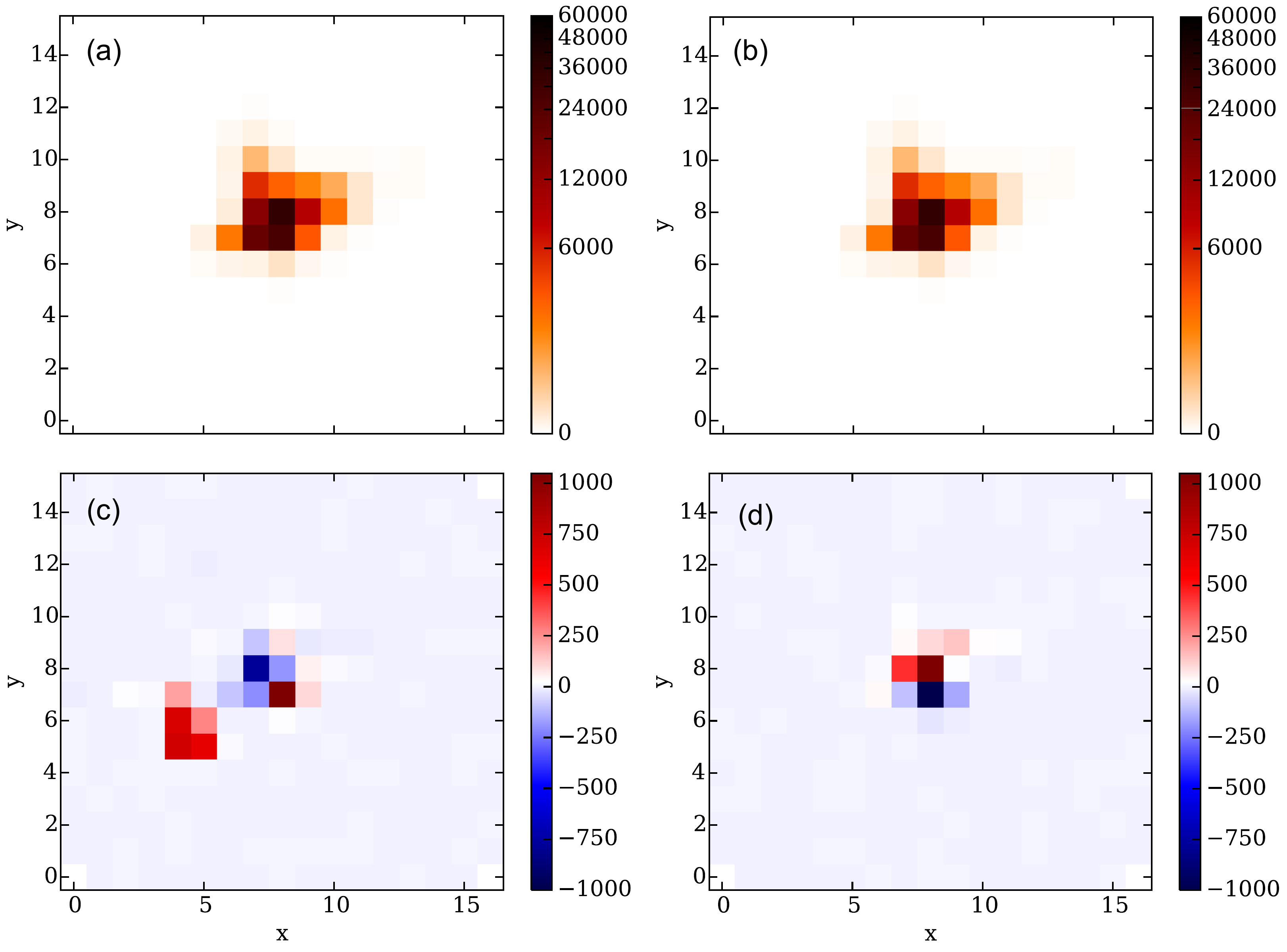}\protect\caption{Cosmic ray detection for \textit{K2} photometry. Panel (a) shows the
background subtracted image of K2-18b for the observation affected
by a cosmic-ray hit (red data point in Figure \ref{fig:Spitzer-light-curve-2-1}).
Panel (c) shows the same frame after subtracting the ``most similar
frame'' in the \textit{K2} time series of K2-18b (Panel (b)). The
cosmic hit is clearly identified in the difference image near x=4.5
and y=5.5. For comparison, Panel (d) shows an equivalent difference
image for an observation not affected by a cosmic ray hit. Small residuals
remain near the center of the PSF, but are not mistaken as cosmic
ray hits by our new cosmic ray detection algorithm (Section \ref{sub:A-Cosmic-Ray})
 \label{fig:Cosmic-ray-identification} }
\end{figure*}

As shown in this work, cosmic ray hits can substantially affect the
astrophysical results derived from \textit{K2} photometry. This is
particular troubling because the effects of cosmic ray hits are generally
too small to result in obvious outliers in the long-cadence, 30-minute
\textit{K2} photometry. As a result, cosmic ray hits have stayed unnoticed
to date, and at least for K2-18b have resulted in inaccurate estimates
of the transit parameters. To address this issue and detect cosmic
ray hits in the \textit{K2} photometry, we introduce a generally applicable
algorithm to efficiently identify cosmic ray hits in the presence
of substantial telescope jitter as we see for \textit{K2}.

We apply the algorithm to the K2-18 data as follows. First, we remove
the sky background from each individual frame by subtracting the median
pixel value outside the PSF of the target star. For each background-subtracted
frame in the \textit{K2} image series, we then find a ``most similar
frame'' by identifying the frame in the \textit{K2} image series
that minimizes the weighted sum of the square differences in x and
y centroid location and telescope roll angle. This most similar frame
will generally appear virtually identical to the original frame because
the PSF falls virtually identical on the detector pixels. The most
similar frame can, therefore, be used to subtract the target star
from the image and isolate potential cosmic ray hits. Finally. we
automatically identify cosmic ray hits by searching for $>10\sigma$
outliers in the time series of each pixel in the difference images. 

As an example, Figure \ref{fig:Cosmic-ray-identification} shows the
detection of the cosmic ray hit near the ingress of the second \textit{K2}
transit observation of K2-18b. Panel (a) shows the background subtracted
image of K2-18b for the cosmic-ray affected observation near the ingress
of the second transit. For comparison, panel (b) shows the frame in
the K2-18 data with the most similar centroid position (BJD=2456875.66).
Panels (a) and (b) appear virtually identical despite the strong color
stretching. The cosmic ray hit near x=4.5 and y=5.5 becomes apparent,
however, in the difference image (Panel (c)). The five pixels near
the cosmic ray hit are identified as outliers with a significance
of $>20\sigma$ in the difference image light curves for these pixels.
For comparison, Panel (d) shows the difference image for a regular
frame without cosmic ray hit. Small residuals ($\sim1\%$) in the
difference image remain due to slight differences in the centroid
position and shape of the PSF. However, the residuals are not mistaken
as cosmic ray hits because the pixel values are within the variances
of the difference image light curves for those pixels.

\subsection{Stellar Spectroscopy}

\begin{figure}[t]
\noindent \centering{}\includegraphics[width=1\columnwidth]{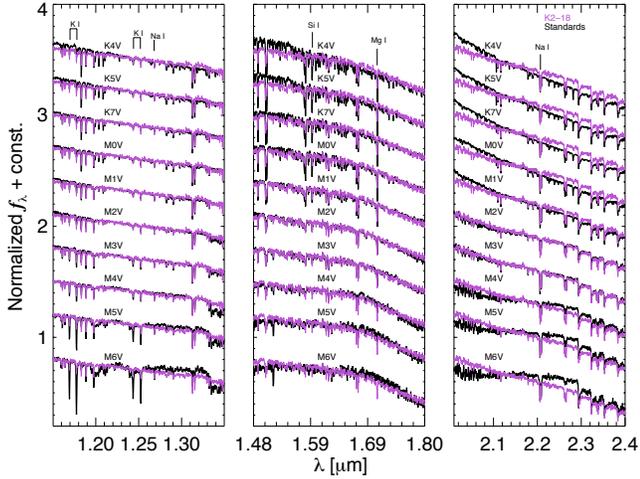}\protect\caption{Calibrated IRTF/SpeX JHK-band spectrum of K2-18 compared to late-type
standards from the IRTF Spectral Library. The spectra are normalize
to the continuum in each band. Across the three bands, the continuum
shape and the strengths of individual absorption features are most
consistent with the M2/M3 standards. This is consistent with the $\mathrm{M2.5\pm0.5}$
spectral type estimated using index based methods. Spectroscopically
derived stellar parameters are listed in Table \ref{tab:Summary-of-Planet}.
\label{fig:Spitzer-light-curve-1-1-1}}
\end{figure}

We observed K2-18 using the near-infrared cross-dispersed spectrograph
(SpeX) on the NASA Infrared Telescope Facility (IRTF) to independently
verify its metallicity ({[}Fe/H{]}), effective temperature ($T_{\mathrm{eff,*}}$),
radius ($R_{*}$), mass ($M_{*}$), and luminosity ($L_{*}$). Following
the procedure described in \citet{crossfield_nearby_2015}, \citet{petigura_two_2015},
and \citet{schlieder_two_2016}, we obtain stellar properties that
are consistent with the stellar properties reported by \citet{montet_stellar_2015}
(Table \ref{tab:Summary-of-Planet}). In short, we observed K2-18
using the short cross-dispersed mode and 0.3x15'' slit providing
simultaneous wavelength coverage from 0.68 to $2.5\,\mathrm{\mu m}$
at a resolution of $R=2000$ (Figure \ref{fig:Spitzer-light-curve-1-1-1}).
The median SNR of our SpeX spectrum is 145 across the JHK bands. Based
on the SpeX spectrum, we estimate the stellar metallicity using empirical
methods based on the spectroscopic indices and equivalent widths calibrated
using M dwarfs that have wide, co-moving FGK companions with well
determined {[}Fe/H{]} \citep{boyajian_stellar_2012,mann_prospecting_2013}.
Similarly, we extract the effective temperature using temperature
sensitive spectroscopic indices in the JHK-bands \citep{mann_spectro-thermometry_2013}
and empirical relations calibrated using nearby, bright M dwarfs \citep{boyajian_stellar_2012}.
We also estimate the spectral type of K2-18 from our SpeX spectrum
using the molecular index based methods of \citet{lepine_spectroscopic_2013}
($\mathrm{TiO_{5}}$, $\mathrm{CaH_{3}}$) and \citet{rojas-ayala_metallicity_2012}
($\mathrm{H\ensuremath{_{2}}O-K_{2}}$). Both methods provide a spectral
type of $\mathrm{M2.5\pm0.5}$. This type is consistent with the derived
stellar parameters and a visual comparison to late-type standards
from the IRTF Spectral Library (Figure 5, \citet{rayner_infrared_2009};
\citet{cushing_infrared_2005}). Finally, we combine the derived $T_{\mathrm{eff,*}}$
and {[}Fe/H{]} and compute the stellar radius and luminosity using
the empirical $T_{\mathrm{eff}}$-{[}Fe/H{]}-$R_{*}$ relation provided
by \citet{mann_how_2015}.

\section{Light Curve Analyses}

\subsection{Spitzer Confirmation of K2-18b\label{sub:Spitzer/IRAC-Instrument-Model}}

\begin{figure*}[t]
\noindent \centering{}\includegraphics[width=1\textwidth]{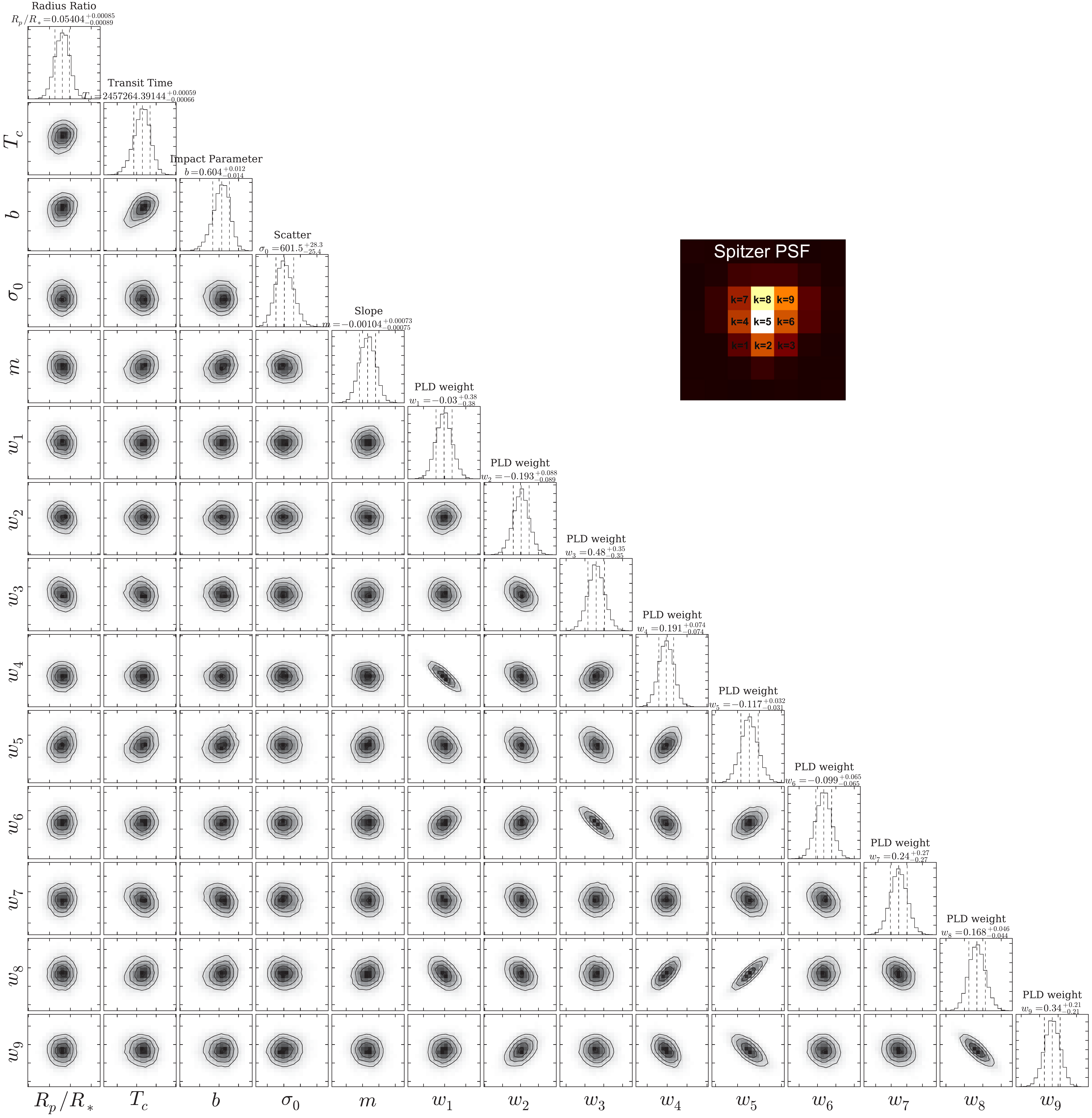}\protect\caption{Pairs plot showing the posterior distribution of the MCMC fitting
parameters for the \textit{Spitzer} light curve fit. The panels on
the diagonal show the marginalized posterior distribution for each
fitting parameter. The 68\% credible interval is marked by vertical
dashed lines and quantified above the panel. The off-diagonal panels
show the two-dimensional marginalized distribution for pairs of parameters,
with the gray shading corresponding to the probability density and
black contours indicating the 68\% and 95\% credible regions. Using
our modified PLD-based \textit{Spitzer} instrument model, we find
that the posterior distributions of all fitting parameters are near-Gaussian,
and that the astrophysical parameters ($R_{P}/R_{*}$, $T_{C}$, $b$)
are virtually uncorrelated with our instrumental parameters ($w_{1}\ldots w_{9},\, m,\,\sigma_{0}$).
The inset at the top right shows a typical image of K2-18's on the
\textit{Spitzer/IRAC} detector, labeling the 3x3 pixels covering the
central region of the PSF. \label{fig:Posteriors-Spitzer}}
\end{figure*}

We analyze the \textit{Spitzer} raw photometry by simultaneously fitting
our \textit{Spitzer/IRAC} instrument model, a transit light curve
model, and a photometric scatter parameter using Markov Chain Monte
Carlo. The entire analysis from raw photometry to the transit parameters
and their uncertainties is performed as a one-step, statistically
consistent Bayesian analysis.

\subsubsection{Spitzer/IRAC Instrument Model\label{sub:Spitzer-Instrument-Model}}

Our \textit{Spitzer/IRAC} instrument model accounts for intrapixel
sensitivity variations and temporal sensitivity changes using a modified
version of the systematics model proposed by \citet{deming_spitzer_2015}.
Our instrument model is 
\begin{equation}
S\left(t_{i}\right)=\frac{\sum_{k=1}^{9}w_{k}D_{k}\left(t_{i}\right)}{\sum_{k=1}^{9}D_{k}\left(t_{i}\right)}+m\cdot t_{i},\label{eq:SpitzerSystematics}
\end{equation}

where the sensitivity function $S\left(t_{i}\right)$ is composed
of the pixel-level decorrelation (PLD) term introduced by \citet{deming_spitzer_2015}
and a linear sensitivity gradient in time. The $D_{k}\left(t_{i}\right)$'s
in the PLD term are the raw counts in the 3x3 pixels, $k=1\ldots9$,
covering the central region of the PSF. In the numerator, these raw
data values are multiplied by the nine time-independent PLD weights,
$\left\{ w_{1}\ldots w_{9}\right\} $, fitted as free parameters in
the light curve analysis. Together with the linear slope $m$, the
instrument model therefore includes 10 free instrument fitting parameters
to capture the intrapixel sensitivity variations and temporal sensitivity
changes. The differences between Equation \ref{eq:SpitzerSystematics}
in this work and Equation 4 in \citet{deming_spitzer_2015} are that
we do not include the offset constant ($h$) and we apply $S\left(t_{i}\right)$
as a multiplicative correction factor corresponding to a variation
in sensitivity rather than an additive term. The log-likelihood function
for fitting the \textit{Spitzer} raw photometry is then 
\begin{equation}
\log L=-\frac{1}{2}\sum_{i=1}^{N}\left(\frac{D\left(t_{i}\right)-S\left(t_{i}\right)\cdot f\left(t_{i}\right)}{\sigma}\right)^{2},\label{eq:likelihoodSpitzer}
\end{equation}

where $f\left(t_{i}\right)$ is the median-normalized flux summed
over all pixels of the target's PSF, $S\left(t_{i}\right)$ is aforementioned
instrument sensitivity, $f\left(t_{i}\right)$ is the model transit
light curve, and $\sigma$ is the photometric scatter parameter simultaneously
fit with the instrument model and transit light curve parameters.

We do not include the constant term $h$ from \citet{deming_spitzer_2015}
because only the relative sizes of the PLD weights carry information
about the intrapixel sensitivity variation. The sum of the PLD weights
$\sum_{k}w_{k}$ uniformly scales the entire light curve up or down,
which is perfectly equivalent to adding a constant $h$. As a result,
if an extra term $h$ was included, it would be perfectly degenerate
with $\sum_{k}w_{k}$ in the fitting, resulting in 100\% degeneracies
between the nine PLD weight and $h$.

We choose to include $S\left(t_{i}\right)$ as a multiplicative correction
factor rather than an additive term as done by \citet{deming_spitzer_2015}
because the multiplicative factor matches more closely the underlying
detector behavior, which is a variation in sensitivity (or quantum
efficiency) across the pixel area. The difference between a multiplicative
factor and a additive term is generally small because $S\left(t_{i}\right)$
is near unity and the multiplication can be approximated by $1+\epsilon$.
However, the inaccuracy due to the negligence of the cross-term $\delta f\cdot\delta S$
can be as high as $0.01\cdot0.005=50\,\textrm{parts-per-million}$
for 1\% transit depth and typical sensitivity variations. We choose
to be on the safe side by introducing a multiplicative correction
factor that correctly captures the cross-term $\delta f\cdot\delta S$.

\subsubsection{Transit Model\label{sub:Transit-Model}}

\begin{figure*}[t]
\noindent \centering{}\includegraphics[width=0.33\textwidth]{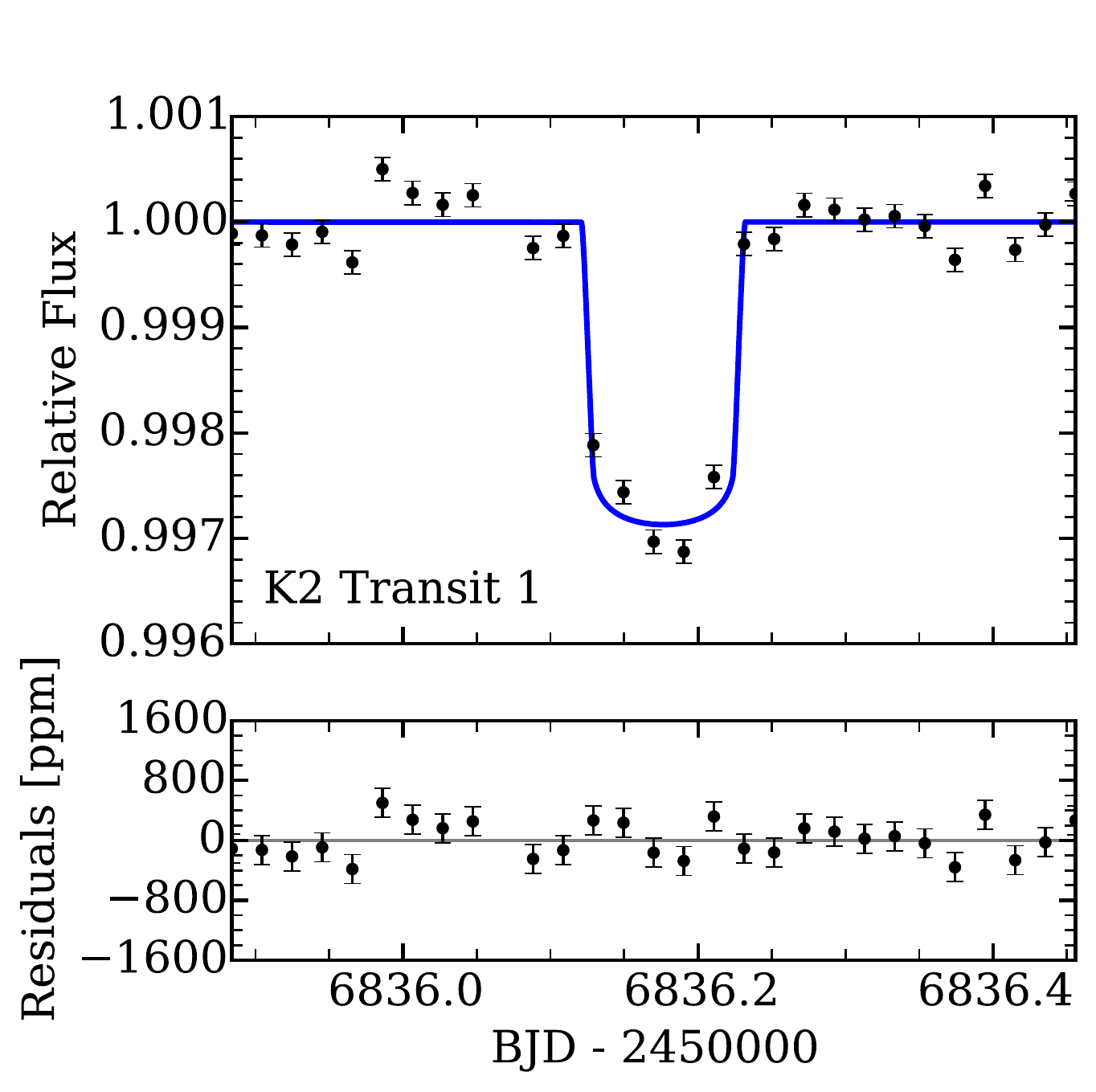}\hfill{}\includegraphics[width=0.33\textwidth]{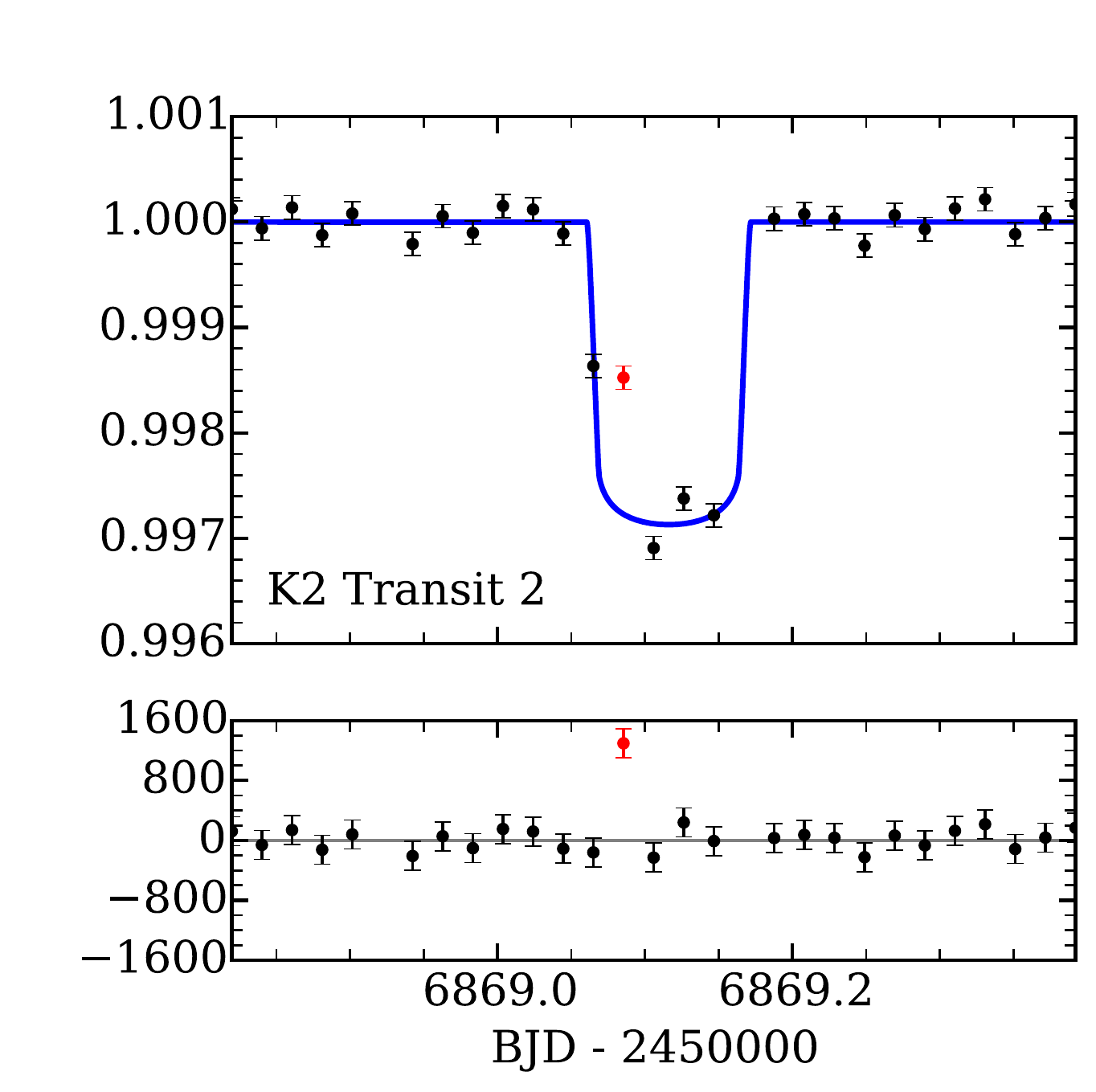}\hfill{}\includegraphics[width=0.33\textwidth]{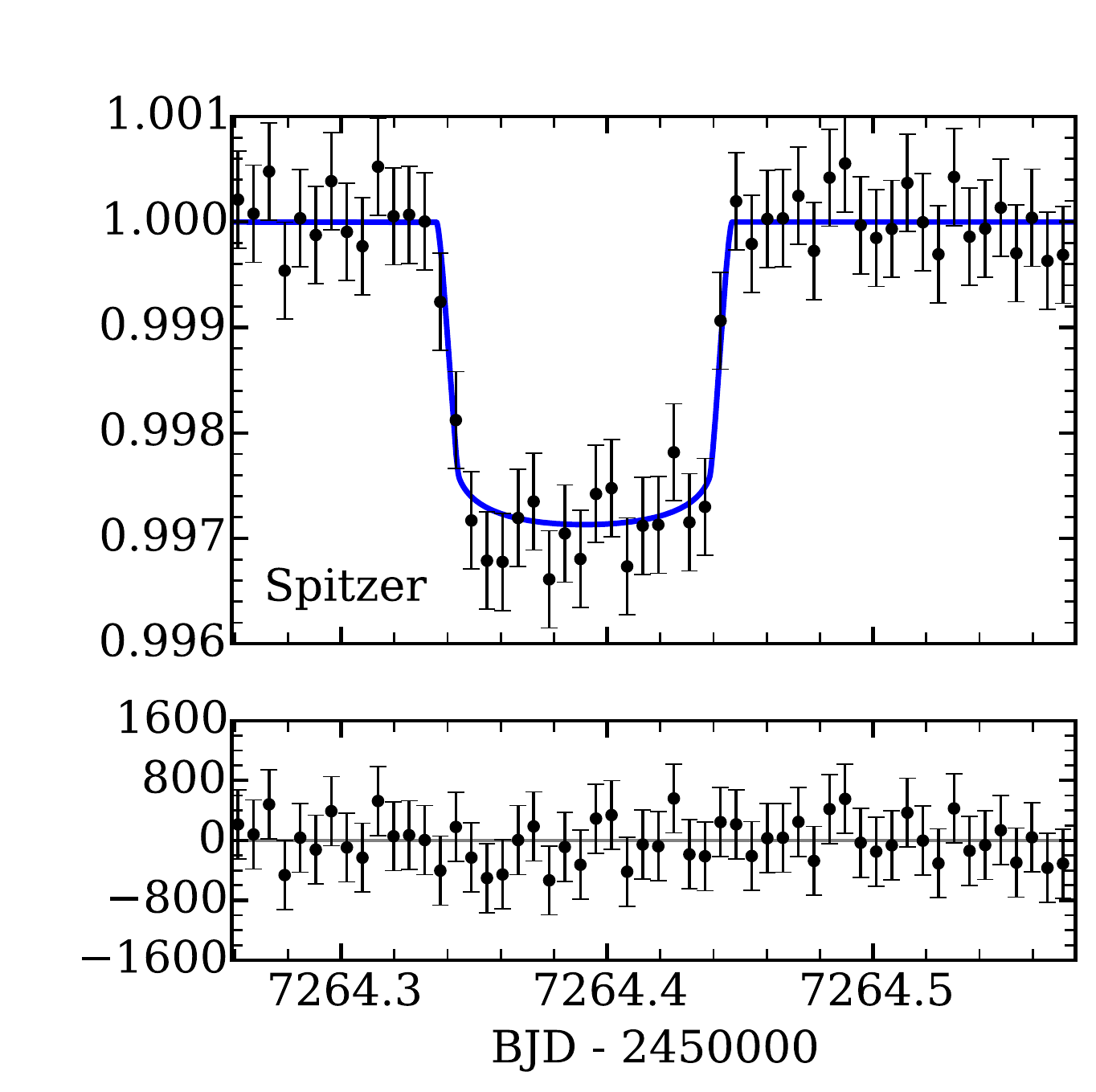}\protect\caption{Joint light curve fit to the \textit{K2} and \textit{Spitzer} observation
of K2-18b. The top panels shows the best fitting model light curve
(blue line), overlaid with the systematics-corrected \textit{K2} data
(left and middle) and \textit{Spitzer} data (right). Residuals from
the light curve fitting are plotted in the bottom panels, with vertical
bars corresponding to the fitted photometric scatter. An outlier data
point affected by a cosmic ray hit near the ingress of the second
\textit{K2} transit is indicated in red and ignored in the light curve
fitting (see also Section \ref{sub:A-Cosmic-Ray}). The high-cadence
\textit{Spitzer} data critically helped identifying the outlier by
precisely constraining the transit duration and transit impact parameter.
\label{fig:Spitzer-light-curve-2-1}}
\end{figure*}

We compute the transit light curve $f\left(t_{i}\right)$ using the
\texttt{Batman} implementation \citep{kreidberg_batman:_2015} of
the Equations derived in \citet{mandel_analytic_2002}. The transit
parameters fitted in our \textit{Spitzer} analysis are the planet-to-star
radius ratio $R_{P}/R_{*}$, the mid-transit time $T_{C}$, and the
impact parameter $b$. We fix the stellar radius and mass to the values
derived based on stellar spectroscopy (Table \ref{tab:Summary-of-Planet})
and assume a circular orbit. We use a quadratic limb darkening profile
with coefficients interpolated from the tables provided by \citet{claret_gravity_2011}
for K2-18's stellar effective temperature and surface gravity ($u_{1}=0.007\pm0.007$,
$u_{2}=-0.191\pm0.005$). These limb darkening coefficients are computed
specifically for the $4.5\,\mathrm{\mu m}$ \textit{Spitzer/IRAC}
Channel 2 bandpass from spherically symmetric Phoenix models \citep[e.g.,][]{hauschildt_nextgen_1999}
using updated opacities. We account for the \ensuremath{\sim}30 minute
cadence of the \textit{K2} observations by numerically integrating
in time.

\subsubsection{MCMC Analysis}

We compute the joint posterior distribution of the instrument and
transit parameters using the \texttt{emcee} package \citep{foreman-mackey_emcee:_2013},
a Python implementation of the Affine Invariant Markov Chain Monte
Carlo (MCMC) Ensemble sampler \citep{goodman_ensemble_2010}. We seed
60 MCMC walkers with initial values widely spread in the prior parameter
space. For convergence, we ensure that the chains for all parameters
are well-mixed as indicated by Gelman-Rubin metrics smaller than 1.02
\citep{gelman_inference_1992}. After an initial burn-in phase, we
generally find good convergence after 3000 to 4000 iterations for
each of the 60 walkers. Since the computational time is not a limiting
factor in this work, we quadruple the number of iterations to obtain
smooth posterior distributions (Figure \ref{fig:Posteriors-Spitzer}).
The final confidence intervals reported in this work are the 15.87\%
and 84.13\% percentiles of each parameters\textquoteright{} posterior
distribution.

\subsubsection{Spitzer Results\label{sub:Spitzer-Results}}

Our \textit{Spitzer} observations robustly reveal a transit event
consistent in transit depth and duration with the super-Earth candidate
K2-18b. The transit event, however, occurs approximately 1.85 hours
($7-\sigma$) before the transit time predicted for K2-18b by \citet{montet_stellar_2015},
which will be discussed further in the following section.

We also find that, using our new modified PLD systematics model, the
final systematics-corrected photometry is near the photon-noise limit
and virtually free of red noise and systematics (Figures \ref{fig:RedNoise}
and \ref{fig:Spitzer-light-curve-1-1}). All posteriors of \textit{Spitzer/IRAC}
systematics parameters are Gaussian-shaped and uncorrelated with the
astrophysical parameters in the transit model, indicating that there
is no dependency of the astrophysical parameters on the instrument
parameters (Figure \ref{fig:Posteriors-Spitzer}).

\subsection{Joint Spitzer/K2 Analysis\label{sub:Joint-Spitzer/K2-Analysis}}

To investigate the source of the $7\,\sigma$ discrepancy between
the predicted and measured transit times, we perform a global analyses
of the \textit{Spitzer} and \textit{K2} data. We directly determine
the transit parameters and their uncertainties from the \textit{K2}
and \textit{Spitzer} raw photometries by simultaneously fitting the
transit light curve model, our Spitzer/IRAC instrument model, and
linear drifts in the photometry of the \textit{K2} transits. We further
assume in this Section \ref{sub:Joint-Spitzer/K2-Analysis} that K2-18b
is orbiting its host star in a Keplerian orbit with a periodic transit
ephemeris and that the transit depths at visible wavelength (\textit{K2})
and IR wavelength (\textit{Spitzer}) are identical. These assumptions
are relaxed in Sections \ref{sub:TTV} and \ref{sub:Visible-IR-Transit-Depth}
and found to be appropriate.

\subsubsection{Instrument Models}

Since we fit the \textit{Spitzer} photometry and two \textit{K2} transits
simultaneously, our fit now includes 14 free parameters for instrument
systematics. As in Section \ref{sub:Spitzer/IRAC-Instrument-Model},
we include nine PLD weights and one linear slope to account for the
intrapixel sensitivity variations and temporal sensitivity changes
in the \textit{Spitzer} photometry (Section \ref{sub:Spitzer-Instrument-Model}).
In addition, two free parameters are included for each of the two
\textit{K2} transits to account for the linear slope and offset in
the \textit{K2} baseline. The log-likelihood function for simultaneously
fitting the \textit{Spitzer} transit and the two \textit{K2} transits
is

\begin{align}\begin{split}\label{eq:Guillot}
	\log L= -\frac{1}{2}\sum_{i=1}^{N_{\mathrm{S}}}\left(\frac{D_{\mathrm{Spitzer}}\left(t_{i}\right)-S\left(t_{i}\right)\cdot f\left(t_{i}\right)}{\sigma_{\mathrm{Spitzer}}}\right)^{2}\\ -\frac{1}{2}\sum_{i=1}^{N_{1}}\left(\frac{D_{\mathrm{K2,1}}\left(t_{i}\right)-\left(m_{1}t_{i}+b_{1}\right)\cdot f\left(t_{i}\right)}{\sigma_{\mathrm{K2}}}\right)^{2}\\ -\frac{1}{2}\sum_{i=1}^{N_{2}}\left(\frac{D_{\mathrm{K2,2}}\left(t_{i}\right)-\left(m_{2}t_{i}+b_{2}\right)\cdot f\left(t_{i}\right)}{\sigma_{\mathrm{K2}}}\right)^{2},
\end{split}\end{align}where $\sigma_{\mathrm{Spitzer}}$ and $\sigma_{\mathrm{K2}}$ are
the fitted photometric scatter values for the \textit{Spitzer} and
\textit{K2} photometry, $N_{s}$, $N_{1}$, and $N_{2}$ are the number
of data points for each of the transit light curves, and $mt_{i}+b$
is the linear baseline for the individual \textit{K2} transit light
curves.
\begin{table*}[t]
\begin{centering}
\begin{tabular}{cccccc}
\toprule 
\multirow{3}{*}{Parameter} & \textit{Spitzer} only & \textit{Spitzer} + \textit{K2} & \textit{Spitzer} + \textit{K2} & \textit{Spitzer} + \textit{K2} & \multirow{3}{*}{Unit}\tabularnewline
 & Keplerian orbit & Keplerian orbit  & Transit timing & Transit depth comp. & \tabularnewline
 &  (Section \ref{sub:Spitzer/IRAC-Instrument-Model}) & (Section \ref{sub:Joint-Spitzer/K2-Analysis}) &  (Section \ref{sub:TTV}) & (Section \ref{sub:Visible-IR-Transit-Depth}) & \tabularnewline
\midrule
\multirow{2}{*}{Radius ratio $R_{P}/R_{*}$} & \multirow{2}{*}{$0.05397_{-0.00089}^{+0.00085}$} & \multirow{2}{*}{$0.05295_{-0.00059}^{+0.00061}$} & \multirow{2}{*}{$0.05303_{-0.00059}^{+0.00059}$} & $0.05205_{-0.00076}^{+0.00077}$ (\textit{Kepler})  & \multirow{2}{*}{1}\tabularnewline
 &  &  &  & $0.05391_{-0.00088}^{+0.00082}$ (\textit{Spitzer}) & \tabularnewline
Impact parameter, $b$ & $0.604_{-0.014}^{+0.012}$ & $0.601_{-0.011}^{+0.013}$ & $0.603_{-0.011}^{+0.011}$ & $0.601_{-0.011}^{+0.012}$ & 1\tabularnewline
 &  &  &  &  & \tabularnewline
\uline{Ephemeris:} &  &  &  &  & \tabularnewline
Mid-transit time, $T_{C}$ & $2457264.39144_{-0.00066}^{+0.00059}$ & $2457264.39131_{-0.00067}^{+0.00060}$ &  & $2457264.39135_{-0.00066}^{+0.00062}$ & BJD\tabularnewline
Orbital period, $P$ & 32.94 (fixed) & $32.939614_{-0.000084}^{+0.000101}$ &  & $32.939622_{-0.000094}^{+0.000099}$ & days\tabularnewline
 &  &  &  &  & \tabularnewline
\uline{Individual transit times:} &  &  &  &  & \tabularnewline
\textit{K2} Transit 1 &  &  & $2456836.1767_{-0.0026}^{+0.0008}$ &  & BJD\tabularnewline
\textit{K2} Transit 2 &  &  & $2456869.11526_{-0.00080}^{+0.00076}$ &  & BJD\tabularnewline
\textit{Spitzer} &  &  & $2457264.39141_{-0.00063}^{+0.00059}$ &  & BJD\tabularnewline
\bottomrule
\end{tabular}
\par\end{centering}

\protect\caption{Transit parameters derived from the \textit{Spitzer} and \textit{K2}
light curves of K2-18b. \label{tab:Transit-parameters-derived} }
\end{table*}
\begin{table}[t]
\begin{centering}
\begin{tabular}{lllc}
\toprule 
Param &  & Units & K2-18b\tabularnewline
\midrule
$T_{0}$ &  & BJD & $2457264.39144_{-0.00066}^{+0.00059}$\tabularnewline
$P$ &  & d & $32.939614_{-0.000084}^{+0.000101}$\tabularnewline
$b$ &  & $-$ & $0.601_{-0.011}^{+0.013}$\tabularnewline
$R_{\mathrm{P}}/R_{*}$ &  & $\%$ & $5.295_{-0.059}^{+0.061}$\tabularnewline
 &  &  & \tabularnewline
$a$ &  & AU & $0.1429_{-0.0065}^{+0.0060}$\tabularnewline
$i$ &  & deg & $89.5785_{-0.0088}^{+0.0079}$\tabularnewline
$R_{\mathrm{P}}$ &  & $R_{\oplus}$ & $2.279_{-0.025}^{+0.026}$\tabularnewline
$S_{\mathrm{inc}}$ &  & W & $1432_{-270}^{+293}$\tabularnewline
$S_{\mathrm{inc}}$ &  & $S_{\oplus}$ & $1.05_{-0.20}^{+0.22}$\tabularnewline
$T_{14}$ &  & min & $159.78_{-1.62}^{+1.40}$\tabularnewline
$T_{23}$ &  & min & $135.24_{-1.96}^{+1.74}$\tabularnewline
 &  &  & \tabularnewline
$R_{*}$ &  & $R_{\odot}$ & $0.411\pm0.038$\tabularnewline
$M_{*}$ &  & $M_{\odot}$ & $0.359\pm0.047$\tabularnewline
$T_{\mathrm{eff,*}}$ &  & K & $3457\pm39$\tabularnewline
{[}Fe/H{]} &  & (dex) & $0.123\pm0.157$\tabularnewline
$\rho_{\mathrm{*,spec}}$ &  & $\mathrm{g\, cm^{-3}}$ & $7.87\pm1.26$\tabularnewline
\bottomrule
\end{tabular}
\par\end{centering}

\protect\caption{Summary of planet and host star properties. \label{tab:Summary-of-Planet}}
\end{table}

\subsubsection{Transit Model}

Our global light curve fit includes a wavelength-independent planet-to-star
radius ratio $p=R_{P}/R_{*}$, the impact parameter $b$, the mid-transit
time $T_{C}$ , and the orbital period $P$. As in Section \ref{sub:Transit-Model},
we fix the stellar radius and stellar mass to the values derived based
on stellar spectroscopy (Table \ref{tab:Summary-of-Planet}) and assume
a circular orbit. For the $4.5\,\mathrm{\mu m}$ \textit{Spitzer/IRAC}
Channel 2 observations, we use the same quadratic limb darkening profile
($u_{1}=0.007\pm0.007$, $u_{2}=-0.191\pm0.005$) as in Section \ref{sub:Transit-Model}.
For the visible \textit{Kepler} bandpass, we derive the quadratic
limb darkening coefficients ourselves using Phoenix models and obtain
$u_{1}=0.153\pm0.004$ and $u_{2}=0.261\pm0.007$.

\subsubsection{Spitzer/K2 Results}

Our joint analysis of the \textit{Spitzer and K2} data reveals that
a previously undetected outlier point in the \textit{K2} photometry
near the ingress of second \textit{K2} transit is the reason for the
1.85 hours ($7\,\sigma$) discrepancy between the transit time predicted
from the \textit{K2} data and the transit time observed by \textit{Spitzer}
(Figure \ref{fig:Spitzer-light-curve-2-1}). Using our newly developed
cosmic ray detection algorithm we find that the outlier is caused
by a cosmic ray hit near the edge of the target star's PSF on the
detector (Figure \ref{fig:Cosmic-ray-identification}). After removal
of the outlier point, we find that a Keplerian orbit with periodic
transit events provide a good joint fit to the \textit{K2} and \textit{Spitzer}
transit light curves (Figure \ref{fig:Spitzer-light-curve-2-1}).
Our new best estimate for the transit ephemeris of K2-18b is $T_{0}=2457264.39131_{-0.00067}^{+0.00060}$
and $P=32.939614_{-0.000084}^{+0.000101}$. 

As for the Spitzer-only fit in Section \ref{sub:Spitzer-Results},
we find that the posteriors of all transit light parameters are Gaussian-shaped
and uncorrelated with the parameters in the systematics model. This
indicates that the derived astrophysical parameters are independent
of the fitted instrument parameters in our joint \textit{Spitzer}/\textit{K2}
fit. We further find that including the \textit{K2} in the fit does
not compromise the excellent noise characteristic of the fit to the
\textit{Spitzer} data (compare Figure \ref{fig:Spitzer-light-curve-1-1}
and \ref{fig:Spitzer-light-curve-2-1}(right panel)).

\subsection{Individual Transit Times\label{sub:TTV}}

\begin{figure}[t]
\noindent \centering{}\includegraphics[width=1\columnwidth]{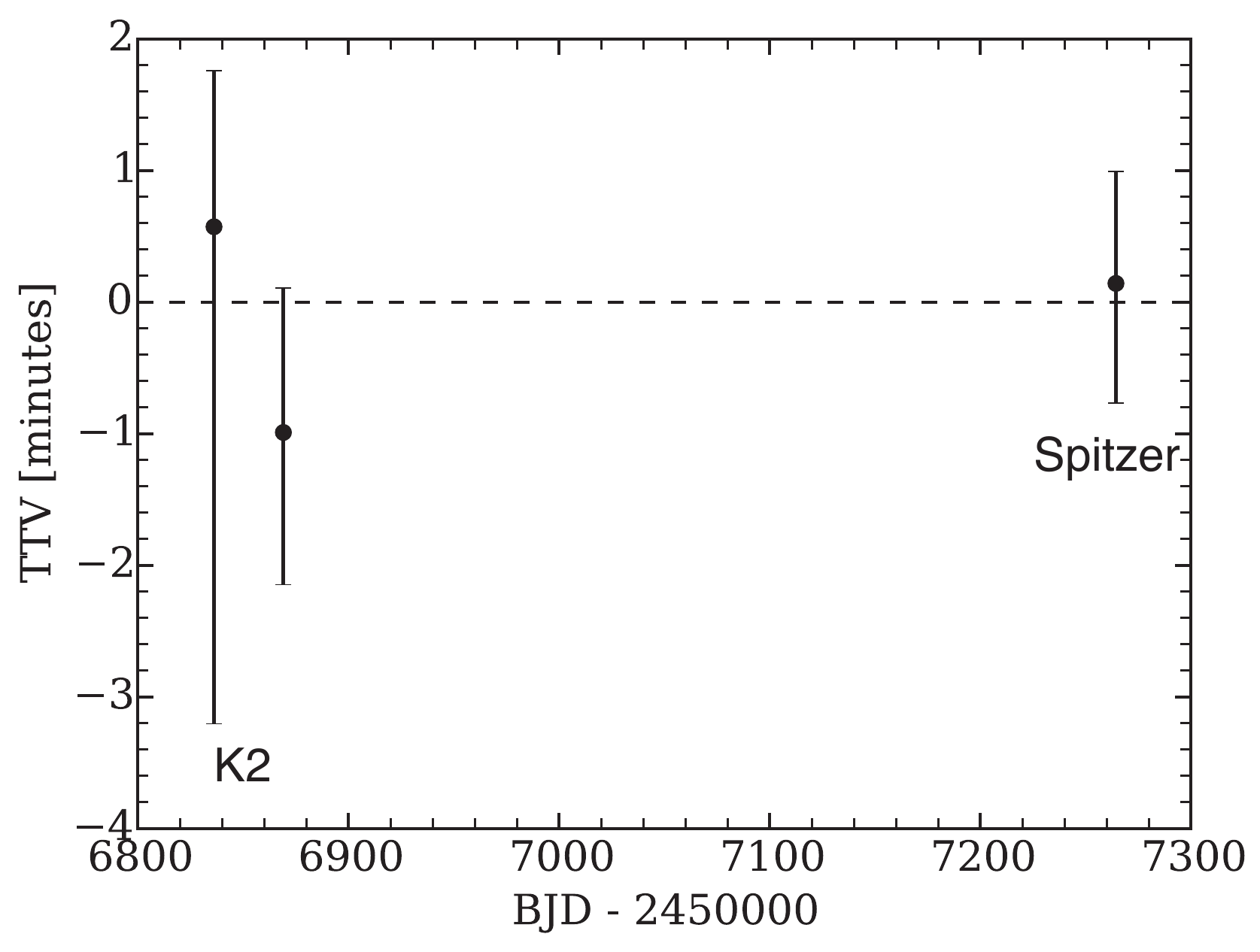}\protect\caption{Transit times of K2-18b relative to the best-fitting linear ephemris
extracted from the global fits to the \textit{K2} and \textit{Spitzer}
data. After removing the outlier in the \textit{K2} photometry, the
data of K2-18b are well explained by Keplerian orbit with linear ephemeris.
\label{fig:TTV}}
\end{figure}
We perform a second global fit to the \textit{K2} and \textit{Spitzer}
data to demonstrate that the initial 1.85 hours discrepancy between
the transit time observed by \textit{Spitzer} (Section \ref{sub:Spitzer/IRAC-Instrument-Model})
and the expected transit time based on Montet al. 2015 is well explained
by the previously undetected cosmic ray hit in the \textit{K2} data.
Our transit timing analysis is identical to the analysis presented
in Section \ref{sub:Joint-Spitzer/K2-Analysis} except that we do
not fit for an average orbital period, but instead parameterize the
mid-transit times of all three individual transits individually (Table
\ref{tab:Transit-parameters-derived}) We remove the discrepant data
point near the ingress of the second \textit{K2} transit (Figure \ref{fig:Spitzer-light-curve-2-1}).
Finally, we probe for deviations from a linear ephemeris derived in
Section \ref{sub:Joint-Spitzer/K2-Analysis} by plotting the differences
between the individually fitted transit times and the calculated transit
times from the best-fitting linear transit ephemeris (Figure \ref{fig:TTV}).

We find that all three measured transit times are fully consistent
with a linear ephemeris to within the timing uncertainties of $1-3$
minutes. We conclude that the \textit{K2} and \textit{Spitzer} data
are well explained by a single planet in a Keplerian orbit. Future
transit observations will be needed to rule transit timing variations
below the $1-3$ minute level or with periods several times greater
than the $430$ days covered by the \textit{K2} and \textit{Spitzer}
observations analyzed here.

\subsection{Transit Depth Comparison between \textit{K2} and \textit{Spitzer}\label{sub:Visible-IR-Transit-Depth}}

\begin{figure}[t]
\noindent \centering{}\includegraphics[width=1\columnwidth]{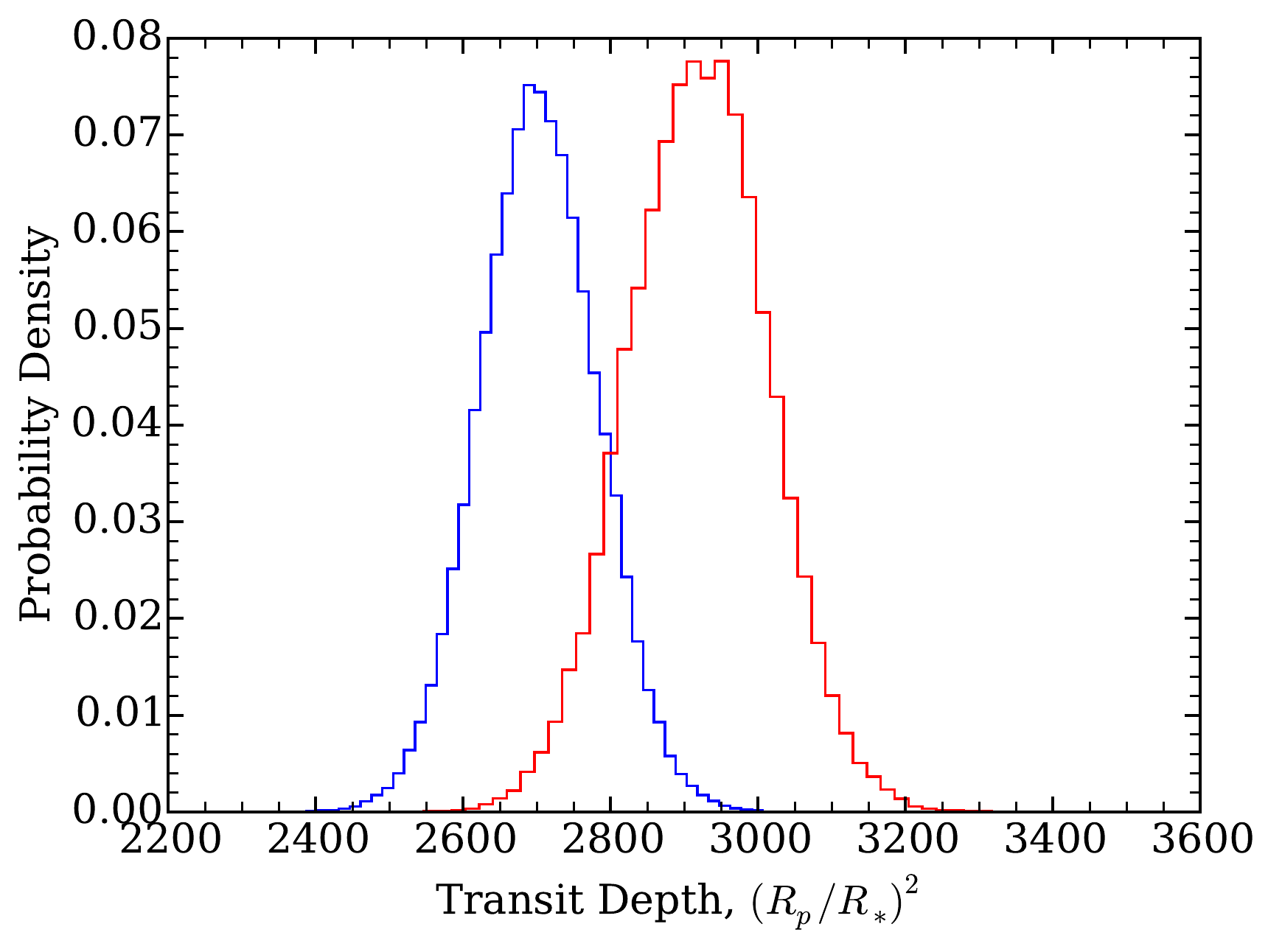}\protect\caption{Marginalized posterior distributions of the transit depths in the
visible-light \textit{K2} bandpass ($0.4-0.7\,\mathrm{\mu m}$, blue)
and in the infrared \textit{Spitzer/IRAC} Channel 2 bandpass ($4-5\,\mathrm{\mu m}$,
red). The fitted transit depths from the Spitzer and \textit{K2} data
are consistent to within approximately $1-\sigma$. \label{fig:Transit-Depth-Comparison}}
\end{figure}
We perform a third global analysis of the \textit{K2} and \textit{Spitzer}
light curves to compare the transit depths in the visible-light \textit{K2}
bandpass ($0.4-0.7\,\mathrm{\mu m}$) and the infrared \textit{Spitzer/IRAC}
bandpass ($4-5\,\mathrm{\mu m}$). Widely different transit depths
at visible and infrared wavelengths could alert us to the potential
presence of a blended star that would affect the inferred planetary
radius measurement \citep{stevenson_deciphering_2014} or even be
the source of a false positive scenario \citep{desert_low_2015}.
Different transit depths could also result from wavelength-dependent
extinction in the atmosphere of K2-18b or the presence of an exosphere
or planetary rings.

We perform the global analysis identical to the one described in Section
\ref{sub:Joint-Spitzer/K2-Analysis}; however, this time we allow
for different transit depths to fit the \textit{K2} and \textit{Spitzer}
transit observations (Table \ref{tab:Transit-parameters-derived}).
We find that the transit depths at visible-light and infrared are
consistent to within the $1\,\sigma$ uncertainties (Figure \ref{fig:Transit-Depth-Comparison}),
ruling out any blended stars or planetary rings that would affect
the transit depth measurement at visible and near-infrared wavelength
by more than $10\%$ (300~ppm). The precision of the transit depth
measurements, however, is currently insufficient to detect gravitationally
bound atmospheres.

\section{Discussion and Conclusions}

\begin{figure}[t]
\noindent \centering{}\includegraphics[width=1\columnwidth]{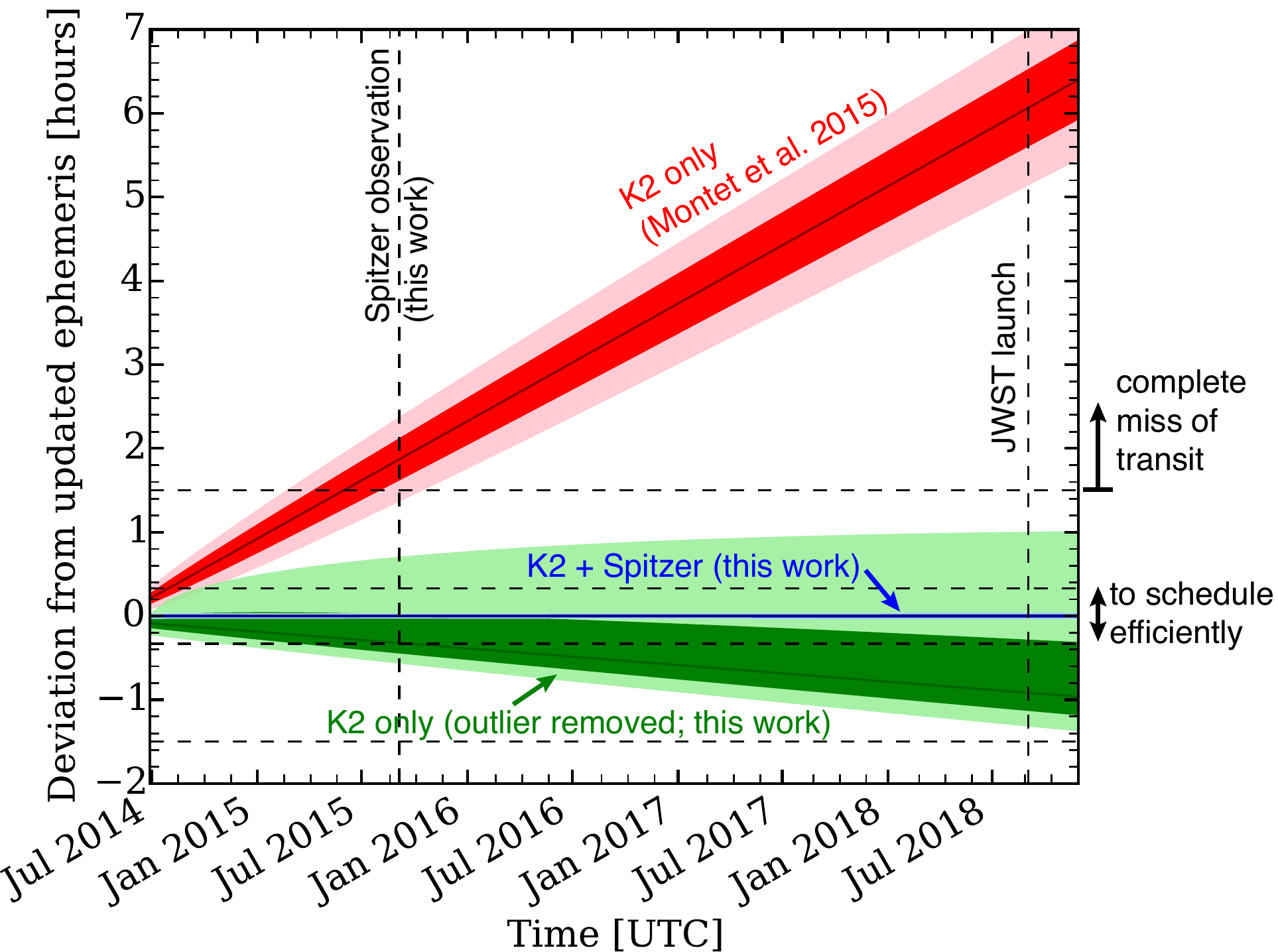}\protect\caption{Deviations from the updated ephemeris. The red region illustrates
the deviation in the predicted mid-transit time between the \textit{K2}-derived
ephemeris reported by Montet et al. 2015 and our updated ephemeris
based on the joint \textit{K2-Spitzer} analysis (blue). Dark and light
red regions correspond to 68\% and 95\% confidence, respectively.
Equivalently the blue regions (within the line thickness) correspond
to the uncertainties of the updated \textit{K2-Spitzer} ephemeris.
Without the immediate Spitzer follow-up, K2-18b would have been for
lost for future atmospheric characterization due to the increasing
deviation in the predicted transit time. For comparison, the green
region indicate the transit timing uncertainty by refitting only the
\textit{K2} data without the outlier data point. The ephemeris based
on only \textit{K2} remains uncertain to 1-2 hours, but the bias is
eliminated by removing the outlier\textit{.} \label{fig:Difference-in-transit}}
\end{figure}
The\textit{ Spitzer Space Telescope} observations presented in this
work confirm the presence of the habitable zone super-Earth K2-18b
by detecting a third transit event with a consistent transit depth
near the predicted transit time. The revealed periodicity of the transit
signal demonstrates that the two transit-like events observed by \textit{K2}
are indeed caused by one planet in a 33-day orbit and are not two
independent events caused by two similarly-sized planets in >50-day
orbits. The periodicity also rules out any scenarios in which one
or both of the identified transit-like events were the result of residual
systematic effects in the corrected \textit{K2} photometry. The photometric
confirmation of K2-18b is critical for future atmospheric studies
because K2-18b is an extremely favorable habitable-zone exoplanet
for transmission spectroscopy with \textit{HST} and \textit{JWST}. 

We also find, however, that the third transit event occurred 1.85
hours ($7-\sigma)$ before the predicted transit time based on the
\textit{K2}-derived ephemeris by \citet{foreman-mackey_systematic_2015}
and \citet{montet_stellar_2015}. Our global analysis of the \textit{K2}
and \textit{Spitzer} data reveals that this 1.85-hour deviation is,
however, not caused by TTVs from another planets in the system, but
is well-explained by a single, previously undetected cosmic ray hit
in the \textit{K2} photometry near the ingress of the second transit.

Our analysis of K2-18b critically reveals that transit ephemerides
of long-period planets based on only two detected transit events can
strongly be affected by individual outlier data points in the \textit{K2}
photometry. A single outlier due to a cosmic ray hit near the ingress
of the second transit biased the ephemeris of K2-18b to a level that
future transit observations could have missed the transit of K2-18b
completely. The deviation in the transit ephemeris would have grown
to 8 hours by the time \textit{JWST} launches (Figure \ref{fig:Difference-in-transit}).
As a result, the first transiting habitable-zone planet amenable to
efficient atmospheric characterization would have been lost for future
spectroscopic transit observations with \textit{HST} or \textit{JWST}
due to the increasing error in its ephemeris estimate. We conclude
that immediate follow-up of prime exoplanet candidates are critical
for long-period planets found by planet search missions such as \textit{K2}
and \textit{TESS}. 

Similarly, the previously undetected outlier in the \textit{K2} photometry
introduced substantial uncertainty in the inference of the planet-to-star
radius ratio. After identifying the cosmic ray hit and removing the
outlier we estimate the planet-to-star radius ratio to $R_{\mathrm{P}}/R_{*}=5.295\%_{-0.059\%}^{+0.061\%}$.
If we ignore our knowledge about the cosmic ray hit and include the
outlier data point in our analysis of the \textit{K2} light curve,
we find the radius ratio uncertainty to be 9 times larger consistent
with $5.13\%{}_{-0.35\%}^{+0.56\%}$ as reported by \citet{foreman-mackey_systematic_2015}.
In this latter case, the wide and asymmetric uncertainties arise because
the outlier data point adversely affects the overall fit to the low-cadence
\textit{K2} photometry. We present an efficient search algorithm to
identify cosmic ray hits in photometry data sets with substantial
telescope pointing jitter to avoid similar problems in future \textit{K2}
or \textit{TESS} light curve analyses.

In the coming years, mass measurement of K2-18b will be critical to
provide an understanding of the nature and bulk composition of K2-18b.
Radial velocity measurements are challenging, however, because the
host stars is faint at visible wavelengths ($V=13.5$) and the expected
radial velocity semi-amplitude is small ($K_{\mathrm{P}}=1-2\,\mathrm{m/s})$.
Still, thanks to the star's brightness in the near-IR ($K=8.9$),
K2-18b may present an ideal target for intensive follow-up with a
number of upcoming NIR radial velocity instrument such as CARMENES
\citep{quirrenbach_carmenes._2012}, SPIRou \citep{artigau_spirou:_2014},
IRD \citep{tamura_infrared_2012}, and CRIRES \citep{kaeufl_crires:_2004}.
In addition, upcoming visible-light radial velocity instruments on
large telescopes like VLT/ESPRESSO \citep{pepe_espresso:_2014}, Keck/SHREK,
and GMT/G-CLEF \citep{szentgyorgyi_gmt-cfa_2012} should also be able
to measure planetary mass of K2-18b in the coming years.

The infrared brightness and small stellar radius of the host star
make K2-18b an extremely favorable candidate for the first detailed
atmospheric characterization of a habitable-zone super-Earth. Given
its radius of $2.27$$R_{\oplus}$, the planet is likely surrounded
by a thick gaseous envelope \citep[e.g.,][]{rogers_most_2015} that
could be amenable to characterization through transit spectroscopy.
Eventually the detectability of the K2-18b's atmosphere will depend
on the mean molecular mass of the atmosphere and presence of high-altitude
clouds \citep{miller-ricci_atmospheric_2009,benneke_how_2013}. In
addition, the range of plausible atmospheric scenarios for K2-18b
also depends on the yet unknown planetary mass and surface gravity.
Little is known about the nature of planets in the habitable zone
around M stars, making K2-18b a unique opportunity to probe chemical
composition and formation history with future follow-up observations.

The \textit{K2} and \textit{Spitzer} analyses presented in this work
were performed using ExoFit, a newly-developed, Python-based light
curve analysis framework. The new framework is highly modular in that
it can jointly fit any number of \textit{Kepler}, \textit{Spitzer},
\textit{HST WFC3}, and/or \textit{HST STIS} transit observations in
a global MCMC analysis with minimum user input. In this work, the
joint analysis of \textit{Spitzer} and \textit{K2} data provides substantial
advantage over individual transit fits because the high cadence \textit{Spitzer}
observations provide exquisite constraints on thetransit duration
that helps fitting the low-cadence \textit{K2} data. For the analysis
of the \textit{Spitzer} observations, we introduce two modifications
to the pixel-level-decorrelation (PLD) approach introduced by \citet{deming_spitzer_2015}.
We find that these changes can provide substantial advantages in the
convergence and uncertainty estimation. With the modifications, the
posterior distribution of all PLD weights in our analyses converge
to Gaussian-shaped posteriors that are uncorrelated with the astrophysical
parameters, providing confidence that the derived transit light curve
parameters are independent of the instrument parameters. Critically,
our corrected \textit{Spitzer} light curve is virtually free of residual
red noise or systematics. The photometric precision of our final Spitzer
light curve is near the Poisson limit.

~

This work is based in part on observations made with the \textit{Spitzer
Space Telescope}, which is operated by the Jet Propulsion Laboratory,
California Institute of Technology under a contract with NASA. Support
for this work was provided by NASA through grants under the HST-GO-13665
program from the STScI and through an award issued by JPL/Caltech.
A.~W.~H. acknowledges support for our \textit{K2} team through a
NASA Astrophysics Data Analysis Program grant. A.~W.~H. and I.~J.~M.~C.
acknowledge support from the \textit{K2} Guest Observer Program.

\textit{Facility}: \textit{Spitzer}, \textit{Kepler}, \textit{K2},
IRTF (SpeX)

\bibliographystyle{apj}
\addcontentsline{toc}{section}{\refname}\bibliography{MyLibrary}

\end{document}